\documentclass[11pt,a4paper]{article}
\usepackage{amsmath,amssymb}
\usepackage{epsfig,graphicx}
\usepackage{cite}

\begin{document}

\title{\bf Plasma induced neutrino spin-flip in a supernova \\ 
and new bounds \\ on the neutrino magnetic moment}

\author{A.~V.~Kuznetsov$^a$\footnote{{\bf e-mail}: avkuzn@uniyar.ac.ru},
N.~V.~Mikheev$^{a}$\footnote{{\bf e-mail}: mikheev@uniyar.ac.ru}
\\
$^a$ \small{\em Yaroslavl State (P.G.~Demidov) University} \\
\small{\em Sovietskaya 14, 150000 Yaroslavl, Russian Federation}
}
\date{}

\maketitle

\begin{abstract}
The neutrino chirality-flip process under the conditions of the supernova core 
is investigated in detail with the plasma polarization effects in the photon 
propagator taken into account in a more complete form than in earlier 
publications. It is shown in part that the contribution of the proton 
fraction of plasma is essential. New upper bounds on the neutrino magnetic 
moment are obtained: 
$\mu_\nu < (0.5 - 1.1) \, \times 10^{-12} \, \mu_{\rm B}$ 
from the limit on the supernova core luminosity for $\nu_R$ emission, 
and $\mu_\nu < (0.4 - 0.6) \, \times 10^{-12} \, \mu_{\rm B}$
from the limit on the averaged time of the left-handed neutrino washing out. 
The best upper bound on the neutrino magnetic moment from SN1987A 
is improved by the factor of 3 to 7.
\end{abstract}
 

\section{Neutrino spin-flip in the supernova core}


\def\D{\mathrm{d}} 
\def\E{\mathrm{e}}
\def\I{\mathrm{i}}

Nonvanishing neutrino magnetic moment leads to various chirality-flipping 
processes where the left-handed neutrinos produced in the stellar interior
become the right-handed ones, i.e. sterile with respect to the weak 
interaction. A considerable interest to the neutrino magnetic moment arised 
after the great event of SN1987A, 
in connection with the modelling of a supernova explosion, where 
gigantic neutrino fluxes define in fact the process energetics. 
It means that such  a microscopic neutrino characteristic, as the neutrino 
magnetic moment, would have a critical influence on macroscopic properties 
of these astrophysical events. Namely, the left-handed neutrinos produced 
inside the supernova core during the collapse, could convert into the 
right-handed neutrinos due to the magnetic moment interaction. 
These sterile neutrinos would 
escape from the core leaving no energy to explain the observed luminosity 
of the supernova. Thus, the upper bound on the neutrino magnetic moment 
can be established. 

This matter was investigated by many authors. 
We will mainly focus on the paper by R. Barbieri and 
R.~N. Mohapatra~\cite{Barbieri:1988} which now looks as the most 
reliable instant constraint on the neutrino magnetic moment
from SN1987A, according to~\cite{RevPartPhys:2006}. 
The authors~\cite{Barbieri:1988} considered the neutrino spin-flip via both 
$\nu_L e^- \to \nu_R e^-$
and $\nu_L p \to \nu_R p$ scattering processes in the inner core of a supernova 
immediately after the collapse. 
Imposing for the $\nu_R$ luminosity $Q_{\nu_R}$ the 
upper limit of $10^{53}$ ergs/s, the authors 
obtained the upper bound on the neutrino magnetic moment:
\begin{eqnarray}
\mu_\nu < (0.2-0.8) \times 10^{-11} \, \mu_{\rm B} \,.
\label{eq:lim_Barbi}
\end{eqnarray}
However, the essential plasma polarization effects in the photon propagator 
were not considered in~\cite{Barbieri:1988}, and the photon dispersion 
was taken in a phenomenolical way, by inserting an {\it ad hoc} thermal mass 
into the vacuum photon propagator. 

A detailed investigation of this question 
was performed in the papers by A. Ayala, J.~C. D'Olivo and 
M. Torres~\cite{Ayala:1999,Ayala:2000}, who 
used the formalism of the thermal field theory to take into account 
the influence of hot dense astrophysical plasma on the photon propagator. 
The upper bound on the neutrino magnetic moment compared with the result 
of the paper~\cite{Barbieri:1988}, was improved in~\cite{Ayala:1999,Ayala:2000} 
in the factor of 2:
\begin{eqnarray}
\mu_\nu < (0.1-0.4) \times 10^{-11} \, \mu_{\rm B} \,.
\label{eq:lim_Ayala}
\end{eqnarray}
However, the authors~\cite{Ayala:1999,Ayala:2000} considered only the 
contribution of plasma electrons, and omitted the proton fraction.
This is despite the fact that the electron and proton contributions to the 
neutrino spin-flip process were evaluated in~\cite{Barbieri:1988} to be 
of the same order.
Thus, the reason exists to reconsider the neutrino spin-flip processes 
in the supernova core more attentively, and this is the subject of this paper. 
Here we confirm in part, that the scattering on plasma protons 
is essential, as well as the scattering on plasma electrons. 


\section{Cherenkov process $\nu_L \to \nu_R \gamma$ and its crossing 
$\nu_L \gamma \to \nu_R$}


Let us start from the Cherenkov process of the photon (plasmon) creation 
by neutrino, $\nu_L \to \nu_R \gamma$, which should be appended 
by the crossed process of the photon absorption 
$\nu_L \gamma \to \nu_R$. 
The Lagrangian of the interaction of a neutrino having a magnetic moment 
$\mu_\nu$ with photons is:
\begin{eqnarray}
{\cal L} = - \frac{\mathrm{i}}{2}  \, \mu_\nu
\left( \bar \nu \, \sigma_{\alpha \beta} \,
\nu \right) F^{\alpha \beta} \,, 
\label{eq:Lagr}
\end{eqnarray}
where $\sigma_{\alpha \beta} = (1/2)\, (\gamma_\alpha \gamma_\beta - 
\gamma_\beta \gamma_\alpha)$, 
$F^{\alpha \beta}$ is the tensor of the photon electromagnetic field.  
For the creation process amplitude one obtains 
\begin{eqnarray}
{\cal M}_{\nu_L \to \nu_R \gamma_\lambda} = \mu_\nu \, j_\alpha \, 
\varepsilon_{(\lambda)}^{\ast \alpha} \,,
\label{eq:ampl}
\end{eqnarray}
where $\varepsilon_{(\lambda)}^{\ast \alpha}$ is the photon 
polarization vector, and $j_\alpha$ is the Fourier transform of the neutrino 
magnetic moment current, 
\begin{eqnarray}
j_\alpha = \left[ \bar \nu_R (p ') \, \sigma_{\mu \alpha} \,
\nu_L (p) \right] q^\mu \,.
\label{eq:nu_current}
\end{eqnarray}
Here, $p^\alpha = (E, {\bf p})$, $p '^\alpha = (E ', {\bf p} ')$ 
and $q^\alpha = (\omega, {\bf k})$ are the four-momenta of the initial and 
final neutrinos and photon, respectively. Note that we use the
signature $(+,-,-,-)$ for the four-metric. 

For the $\nu_L \to \nu_R$ conversion width one obtains by the standard way:
\begin{eqnarray}
\Gamma^{\rm tot}_{\nu_L \to \nu_R} &=& 
\Gamma_{\nu_L \to \nu_R \gamma} +
\Gamma_{\nu_L \gamma \to \nu_R}
= \frac{\mu_\nu^2}{16 \, \pi^2 E} \int j_\alpha \, j^\ast_\beta \,
\sum\limits_{\lambda = t, \ell}  
\varepsilon_{(\lambda)}^{\ast \alpha} \, \varepsilon_{(\lambda)}^\beta 
\, Z_\gamma^{(\lambda)}
\, \frac{\D^3 p'}{E'}
\nonumber\\[2mm]
&\times& \left\{
\frac{\delta (E - E' - \omega)}{2 \, \omega} 
\left[ 1 + f_\gamma (\omega) \right] +
\frac{\delta (E - E' + \omega)}{2 \, \omega} \, 
f_\gamma (\omega) 
\right\} ,
\label{eq:width}
\end{eqnarray}
where 
$\lambda = t, \ell$ mean transversal and longitudinal photon polarizations, 
$f_\gamma (\omega) = \left(\E^{\omega/T} - 1\right)^{-1}$ is the 
Bose---Einstein photon distribution function,
and 
$Z_\gamma^{(\lambda)} = (1 - \partial \Pi_{(\lambda)}/\partial \omega^2)^{-1}$ is the 
photon wave-function renormalization. 
The functions $\Pi_{(\lambda)}$, defining the photon dispersion law: 
\begin{eqnarray}
\omega^2 - {\bf k}^2 - \Pi_{(\lambda)} (\omega, {\bf k}) = 0 \,,
\label{eq:disp}
\end{eqnarray}
are the eigenvalues of the photon polarization tensor 
$\Pi_{\alpha \beta}$, 
\begin{eqnarray}
\Pi_{\alpha \beta} \, \varepsilon_{(\lambda)}^\beta 
= \Pi_{(\lambda)} \, \varepsilon_{(\lambda)\alpha} \,,
\label{eq:Pi_tensor}
\end{eqnarray}
and can be found e.g. in~\cite{Braaten:1993}. 

The width $\Gamma^{\rm tot}_{\nu_L \to \nu_R}$ can be rewritten 
to another form. Let us introduce the energy transferred from neutrino:
$ E - E' = q_0$, which is expressed via the photon energy 
$\omega (k)$ as $q_0 = \pm \omega (k)$. Then 
$\delta$-functions in Eq.~(\ref{eq:width}) can be rewritten, 
\begin{eqnarray}
&&\frac{\delta \left(q_0 - \omega (k) \right)}{2 \,\omega (k)} =  
\delta \left(q_0^2 - \omega^2 (k) \right) \, \theta (q_0), 
\nonumber\\[2mm]
&&\frac{\delta \left(q_0 + \omega (k) \right)}{2 \,\omega (k)} =  
\delta \left(q_0^2 - \omega^2 (k) \right) \, \theta (-q_0). 
\label{eq:delta1}
\end{eqnarray}
Transforming the $\delta$-function to have the 
dispersion law in the argument: 
\begin{eqnarray}
\delta \left(q_0^2 - \omega^2 (k) \right) = 
\left[ Z_\gamma^{(\lambda)} \right]^{-1} \, 
\delta \left(q^2 - \Pi_{(\lambda)} (q) \right) ,
\label{eq:delta2}
\end{eqnarray}
one can see that the renormalization factor 
$Z_\gamma^{(\lambda)}$ is cancelled in the conversion width~(\ref{eq:width}).


\section{The rate of creation of the right-handed neutrino} 


Instead of $\Gamma^{\rm tot}_{\nu_L \to \nu_R}$, 
the physical value we should be more interested in, is e.g. the right-handed 
neutrino flux, integrated over the initial left-handed neutrino states, 
i.e. the number of right-handed neutrinos emitted per unit time:
\begin{eqnarray}
N_{\nu_R} = \int \, \D n_{\nu_L} \, \Gamma^{\rm tot}_{\nu_L \to \nu_R}
= \int \, \frac{\D^3 p \, V}{(2 \, \pi)^3} \, f_{\nu_L} (E) \, 
\Gamma^{\rm tot}_{\nu_L \to \nu_R} \,, 
\label{eq:N_nu}
\end{eqnarray}
where $f_{\nu_L} (E) 
= \left(\E^{(E - \tilde \mu_\nu)/T} + 1\right)^{-1}$ is 
the left-handed neutrino distribution function with the chemical potential 
$\tilde \mu_\nu$. 
There exists even more convenient value, the rate of creation of the 
right-handed neutrino $\nu_R$, $\Gamma_{\nu_R} (E')$, with the fixed energy 
$E'$ by all the left-handed neutrinos: 
\begin{eqnarray}
\Gamma_{\nu_R} (E') = \frac{\D \, N_{\nu_R}}{\D \, n_{\nu_R}}\,, 
\quad 
\D \, n_{\nu_R} = \frac{\D^3 p' \, V}{(2 \, \pi)^3} \,.
\label{eq:Gamma1}
\end{eqnarray}
Given $\Gamma_{\nu_R} (E')$, one can calculate both 
the right-handed neutrino flux and the right-handed neutrino luminosity.

The function $\Gamma_{\nu_R} (E')$ takes the form:
\begin{eqnarray}
\Gamma_{\nu_R} (E') 
&=& \frac{\mu_\nu^2}{16 \, \pi^2 \, E'} \, \int \, \frac{\D^3 p}{E} \, 
f_{\nu_L} (E) \, 
j_\alpha \, j^\ast_\beta \, \sum\limits_{\lambda = t, \ell} \, 
\varepsilon_{(\lambda)}^{\ast \alpha} \, \varepsilon_{(\lambda)}^\beta \, 
\delta \left(q^2 - \Pi_{(\lambda)} (q) \right) 
\nonumber\\[2mm]
&\times& 
\left\{
\left[ 1 + f_\gamma (q_0) \right] \theta (q_0) +
f_\gamma (-q_0) \, \theta (-q_0)
\right\} ,
\label{eq:Gamma3}
\end{eqnarray}
and can be easily calculated when the function 
$\Pi_{(\lambda)} (q)$ is real.
However, it has, in general, an imaginary part.
It means, that the photon is unstable in plasma. 

Nevertheless, there could be a possibility to move forward, if 
one would use instead of the $\delta$-function its 
generalization. We suppose the generalization of the Breit---Wigner type, 
with the retarded functions $\Pi_{(\lambda)} (q)$ :
\begin{eqnarray}
\delta \left(q^2 - \Pi_{(\lambda)} (q) \right) \; \Rightarrow \;
\frac{1}{\pi} \; \frac{- \mathrm{Im} \,\Pi_{(\lambda)} \; \mathrm{sign} (q_0) \, \epsilon_{\lambda}}
{\left(q^2 - \mathrm{Re} \,\Pi_{(\lambda)} \right)^2 + 
\left(\mathrm{Im} \,\Pi_{(\lambda)} \right)^2} \,,
\label{eq:Breit-Wigner}
\end{eqnarray}
where $\epsilon_{\lambda} = + 1$ 
for $\lambda = t$
and $\epsilon_{\lambda} = - 1$ 
for $\lambda = \ell$.

Taking into account that $f_\gamma (-q_0) = 
- \left[ 1 + f_\gamma (q_0) \right] $, one obtains 
\begin{eqnarray}
\mathrm{sign} (q_0) \, \left\{
\left[ 1 + f_\gamma (q_0) \right] \theta (q_0) +
f_\gamma (-q_0) \, \theta (-q_0)
\right\} = 1 + f_\gamma (q_0) \,,
\nonumber
\end{eqnarray}
and the rate of creation of the 
right-handed neutrino takes the form:
\begin{eqnarray}
\Gamma_{\nu_R} (E') 
&=& \frac{\mu_\nu^2}{16 \, \pi^3 \, E'} \, \int \, \frac{\D^3 p}{E} \, f_{\nu_L} (E) \,
\left[ 1 + f_\gamma (q_0) \right] \, j_\alpha \, j^\ast_\beta 
\nonumber\\[2mm]
&\times& 
\sum\limits_{\lambda = t, \ell} \, 
\frac{\rho^{\alpha \beta}_{(\lambda)} 
\left(- \mathrm{Im} \,\Pi_{(\lambda)} \right) \, \epsilon_{\lambda} }
{\left(q^2 - \mathrm{Re} \,\Pi_{(\lambda)} \right)^2 + 
\left(\mathrm{Im} \,\Pi_{(\lambda)} \right)^2} \,,
\label{eq:Gamma4}
\end{eqnarray}
where the polarization density matrices 
for the transversal and longitudinal photons are introduced:
\begin{eqnarray}
\rho^{\alpha \beta}_{(t)} = \sum\limits_{\lambda = t_1, t_2} \, 
\varepsilon_{(\lambda)}^{\ast \alpha} \, \varepsilon_{(\lambda)}^\beta = 
- \left( g^{\alpha \beta} - \frac{q^\alpha \, q^\beta}{q^2} 
 - \frac{\ell^\alpha \, \ell^\beta}{\ell^2} \right) ,
\nonumber\\[2mm]
\rho^{\alpha \beta}_{(\ell)} = 
- \, \varepsilon_{(\ell)}^{\ast \alpha} \, \varepsilon_{(\ell)}^\beta = 
 - \, \frac{\ell^\alpha \, \ell^\beta}{\ell^2} \,, 
\qquad \ell_\alpha = q_\alpha \, (u q) - u_\alpha \, q^2 \,,
\label{eq:density_matrices}
\end{eqnarray}
$u_\alpha$ is the four-vector of the plasma velocity.  

The structures appeared in the function $\Gamma_{\nu_R} (E')$ are 
called the photon spectral density functions:
\begin{eqnarray}
\varrho_{(\lambda)} = 
\frac{2 \left(- \mathrm{Im} \,\Pi_{(\lambda)} \right)}
{\left(q^2 - \mathrm{Re} \,\Pi_{(\lambda)} \right)^2 + 
\left(\mathrm{Im} \,\Pi_{(\lambda)} \right)^2} \,,
\label{eq:spect_fun}
\end{eqnarray}
Changing the integration variables $\D^3 p \to \D q_0 \, \D k$, 
one obtains after calculation:
\begin{eqnarray}
\Gamma_{\nu_R} (E') &=& \frac{\mu_\nu^2}{16\, \pi^2 \, E'^2} \; 
\int\limits_{-E '}^\infty \, \mathrm{d} q_0 \, 
\int\limits_{|q_0|}^{2 E ' + q_0} \, k^3 \, \mathrm{d} k \, 
f_\nu (E' + q_0) \left[ 1 + f_\gamma (q_0) \right] \, (2 E' + q_0)^2 \, 
\nonumber\\[2mm]
&\times& 
\left[ 1 - \left( \frac{q_0}{k} \right)^2 \right]^2
\left[ 
\left( 1 - \frac{k^2}{(2 E' + q_0)^2} \right)\, \varrho_{(t)} 
- \varrho_{(\ell)}  
\right] .
\label{eq:Gamma5}
\end{eqnarray}
%


\section{Neutrino interaction with background}

\begin{figure}[htb]
\centering
\includegraphics[width=0.2\textwidth]{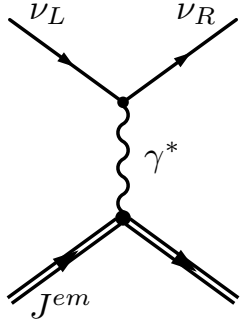}
\caption{The Feynman diagram for the neutrino spin-flip scattering via the 
intermediate plasmon $\gamma^*$ on the plasma electromagnetic current $J^{em}$.}
\label{fig:inter_plasmon}
\end{figure}

Thus, a ``photon'' considered in the neutrino chirality flip processes 
$\nu_L \to \nu_R \gamma$ and $\nu_L \gamma \to \nu_R$, 
obviously can not be treated as a real photon. 
It would be more self-consistent to consider 
the vertex $\nu_L \nu_R \gamma^\ast$ in the neutrino scattering via the 
intermediate virtual plasmon $\gamma^\ast$ on the plasma 
electromagnetic 
current presented by electrons, $\nu_L e^- \to \nu_R e^-$,
protons, $\nu_L p \to \nu_R p$, etc., see the Feynman diagram in 
Fig.~\ref{fig:inter_plasmon}. 
Here, $J^{em}$ is an electromagnetic current in the general 
sense, formed by different components of the medium, i.e. free 
electrons and positrons, free ions, neutral atoms, etc. 

The technics of calculations of the neutrino spin-flip rate is rather standard.
The only principal point is to use the photon propagator 
$G^{\alpha \beta} (q)$ with taking account 
of the plasma polarization effects. We take it in the form: 
\begin{eqnarray}
G^{\alpha \beta} (q) = 
\frac{\I \,\varrho^{\alpha \beta}_{(t)}}{q^2 - \Pi_{(t)}} +
\frac{\I \,\varrho^{\alpha \beta}_{(\ell)}}{q^2 - \Pi_{(\ell)}} \,,
\label{eq:propagator}
\end{eqnarray}
which has no ambiguity when the functions $\Pi_{(t,\ell)}$ are real. 
Our generalization to the case of complex functions is based on using the 
same form of the propagator with the retarded functions 
$\Pi_{(t,\ell)}$. 

Integrating the amplitude squared of the process, described by the Feynman 
diagram of Fig.~\ref{fig:inter_plasmon},
over the states of particles forming 
the electromagnetic current and over the states of the initial 
left-handed neutrinos, we obtain just the same formula~(\ref{eq:Gamma5})
for the rate $\Gamma_{\nu_R} (E')$ of creation of the right-handed 
neutrino with the fixed energy $E'$.

There is also such a subtle effect as the additional energy $W$ 
acquired by a left-handed neutrino in plasma. With this effect, the general 
expression for the rate of creation of the right-handed neutrino is:
\begin{eqnarray}
\Gamma_{\nu_R} (E') &=& \frac{\mu_\nu^2}{16\, \pi^2 \, E'^2} \; 
\int\limits_{-E '}^\infty \, \mathrm{d} q_0 \, 
\int\limits_{|q_0|}^{2 E ' + q_0} \, \frac{\D k}{k} \, 
f_\nu (E' + q_0) \left[ 1 + f_\gamma (q_0) \right] \, (2 E' + q_0)^2 \, q^4
\nonumber\\[2mm]
&\times& 
\left\{ 
\left( 1 - \frac{k^2}{(2 E' + q_0)^2} \right) 
\left[ 1 - \frac{2 q_0 W}{q^2} + \frac{8 E' (E' + q_0) W^2}
{q^4 \left[ (2 E' + q_0)^2/k^2 - 1  \right]} \right] 
\varrho_{(t)} (q_0, k)
\right.
\nonumber\\[2mm]
&-& \left.
\left( 1 - \frac{2 q_0 W}{q^2} \right) \varrho_{(\ell)} (q_0, k) 
\right\} ,
\label{eq:Gamma6}
\end{eqnarray}
where $q^2 = q_0^2 - k^2 $ .

We note that our result is in agreement, to the notations, with the rate 
obtained 
by P. Elmfors et al.~\cite{Elmfors:1997} from the retarded self-energy 
operator of the right-handed neutrino.  
However, extracting from our general expression the electron contribution 
only, we obtain the result which is larger by the factor of 2 
than the corresponding formula in the papers by A. Ayala 
et al.~\cite{Ayala:1999,Ayala:2000}. 
It can be seen that an error was made there just in the first formula 
defining the production rate $\Gamma$ of a right-handed neutrino. 

Our formula having the most general form, can be used for neutrino-photon 
processes in any optically active medium. We only need to identify 
the photon spectral density functions $\varrho_{(\lambda)}$. 
For example, in the medium where $\mathrm{Im} \,\Pi_{(t)} \to 0$ 
in the space-like region $q^2 < 0$ corresponding to the refractive 
index values $n > 1$, the spectral density function is transformed to 
$\delta$-function, and we reproduce the result of the paper by 
W. Grimus and H. Neufeld~\cite{Grimus:1993} devoted to the study of the 
Cherenkov radiation of transversal photons by neutrinos. 

If one formally takes the limit 
$\mathrm{Im} \,\Pi_{(\ell)} \to 0$, 
the result obtained by S. Mohanty and S. Sahu~\cite{Mohanty:1997} can be 
reproduced, namely, the width of the Cherenkov radiation and absorption of 
longitudinal photons by neutrinos in the space-like region 
$q^2 < 0$. However, the 
limit $\mathrm{Im} \,\Pi_{(\ell)} \to 0$ itself is unphysical in the real 
astrophysical plasma conditions considered by those authors and leads to 
the strong overestimation of a result.

\begin{figure}[htb]
\centering
\includegraphics[width=0.95\textwidth]{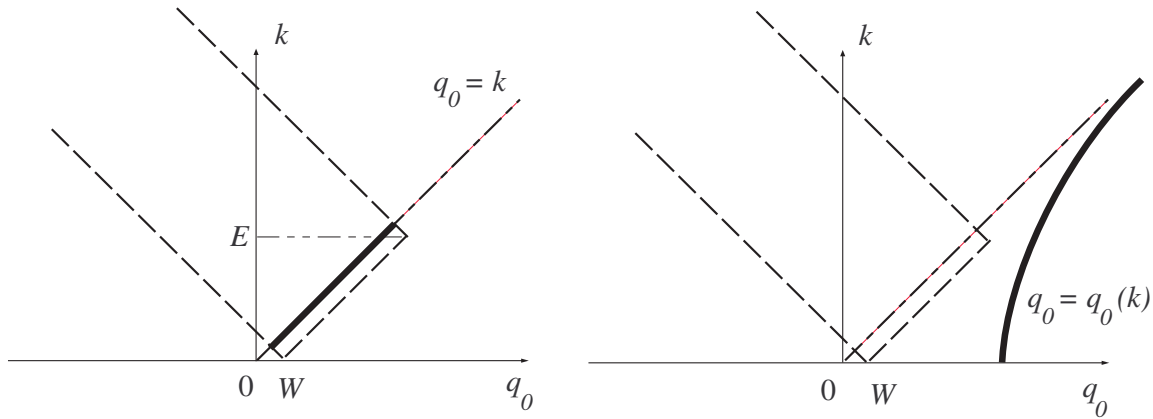}
\caption{
The region of integration for the width $\Gamma^{\rm tot}_{\nu_L \to \nu_R}$ 
with the fixed initial neutrino energy $E$ in the unphysical case when 
the photon dispersion in medium was ignored at all 
(bold line in the left plot). 
With the photon dispersion in account (bold line in the right plot)
the threshold neutrino energy $E_{\mathrm{min}}$ exists for coming of the 
dispersion curve into the allowed kinematical region.
}
\label{fig:disp}
\end{figure}

One more unphysical case was considered in the papers by 
A.~Studenikin et al.~\cite{0611100}, where the additional energy $W$ 
acquired by a left-handed neutrino in plasma was taken, but 
the photon dispersion in medium was ignored at all. 
The region 
of integration for the width $\Gamma^{\rm tot}_{\nu_L \to \nu_R}$ 
with the fixed initial neutrino energy $E$ 
was the vacuum dispersion line $q_0 = k$, see Fig.~\ref{fig:disp}, left plot. 

However, the photon dispersion in plasma is not the vacuum one, 
see Fig.~\ref{fig:disp}, right plot.  
For the fixed plasma parameters, the threshold neutrino energy 
$E_{\mathrm{min}}$ exists for coming of the dispersion curve into 
the allowed kinematical region, as was shown in our 
papers~\cite{Kuznetsov:2006,Kuznetsov:2007}


\section{Right-handed neutrino luminosity}
 

As it was mentioned above, an analysis of the neutrino chirality flip 
process has to be performed with taking account of the neutrino scattering 
off various plasma components: electrons, protons, free ions, etc.
One can consider the contribution of the neutrino scattering 
off electrons into the right-handed neutrino production rate.
This means that one should take into account the electron contribution only 
into the function $\mathrm{Im} \,\Pi_{(\lambda)}$ in the numerator of the 
spectral density function~(\ref{eq:spect_fun}). 
It should be stressed however, that the functions 
$\mathrm{Re} \,\Pi_{(\lambda)}$ and $\mathrm{Im} \,\Pi_{(\lambda)}$ 
in the denominator of Eq.~(\ref{eq:spect_fun}) contain the contributions 
of all plasma components. At this point our result for the neutrino scattering 
off electrons differs from the result 
by A. Ayala et al.~\cite{Ayala:1999,Ayala:2000}, where the electron 
contribution only was taken both in the numerator and in the denominator of the 
plasmon spectral densities.  

\begin{figure}[htb]
\centering
\includegraphics[width=0.8\textwidth]{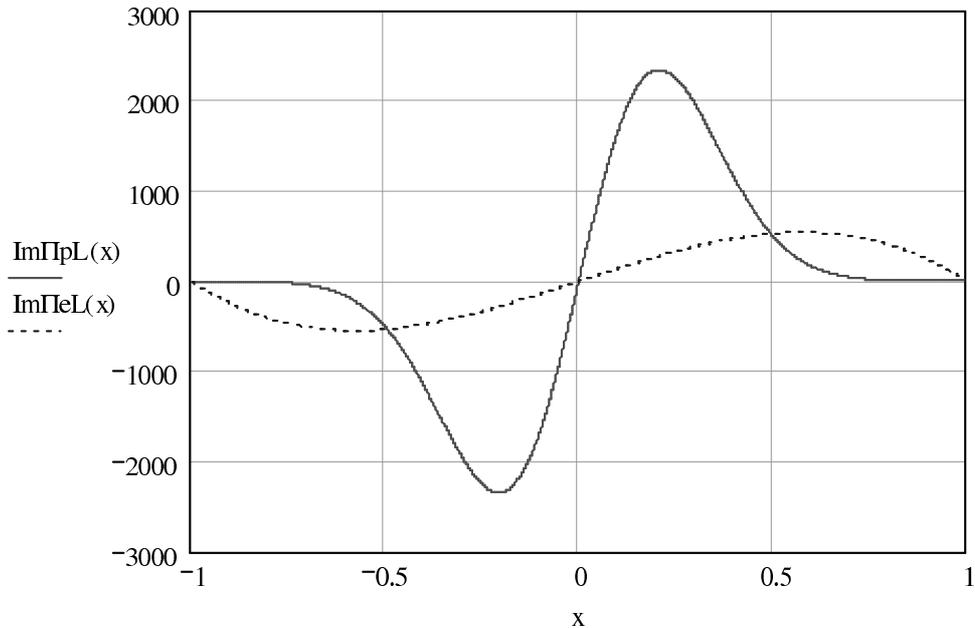}
\caption{Proton (solid line) and electron (dotted line) contributions 
to the imaginary part of $\Pi_{\ell}$.}
\label{fig:Im_Pi_ell}
\end{figure}

\begin{figure}[htb]
\centering
\includegraphics[width=0.8\textwidth]{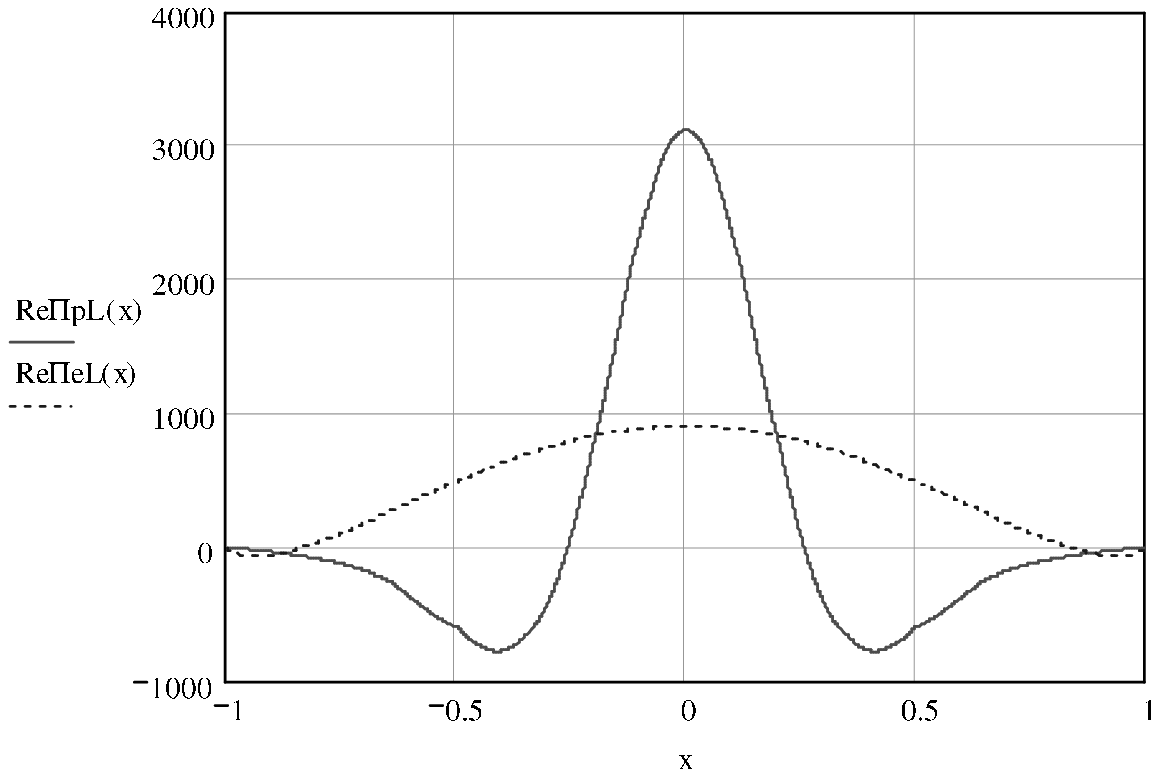}
\caption{Proton (solid line) and electron (dotted line) contributions 
to the real part of $\Pi_{\ell}$.}
\label{fig:Re_Pi_ell}
\end{figure}

In the Figs.~\ref{fig:Im_Pi_ell} and ~\ref{fig:Re_Pi_ell} 
we illustrate the importance of taking into account 
the proton contribution into the eigenvalue $\Pi_{\ell}$ for the 
longitudinal plasmon. 

%

\begin{figure}[htb]
\centering
\includegraphics[width=0.8\textwidth]{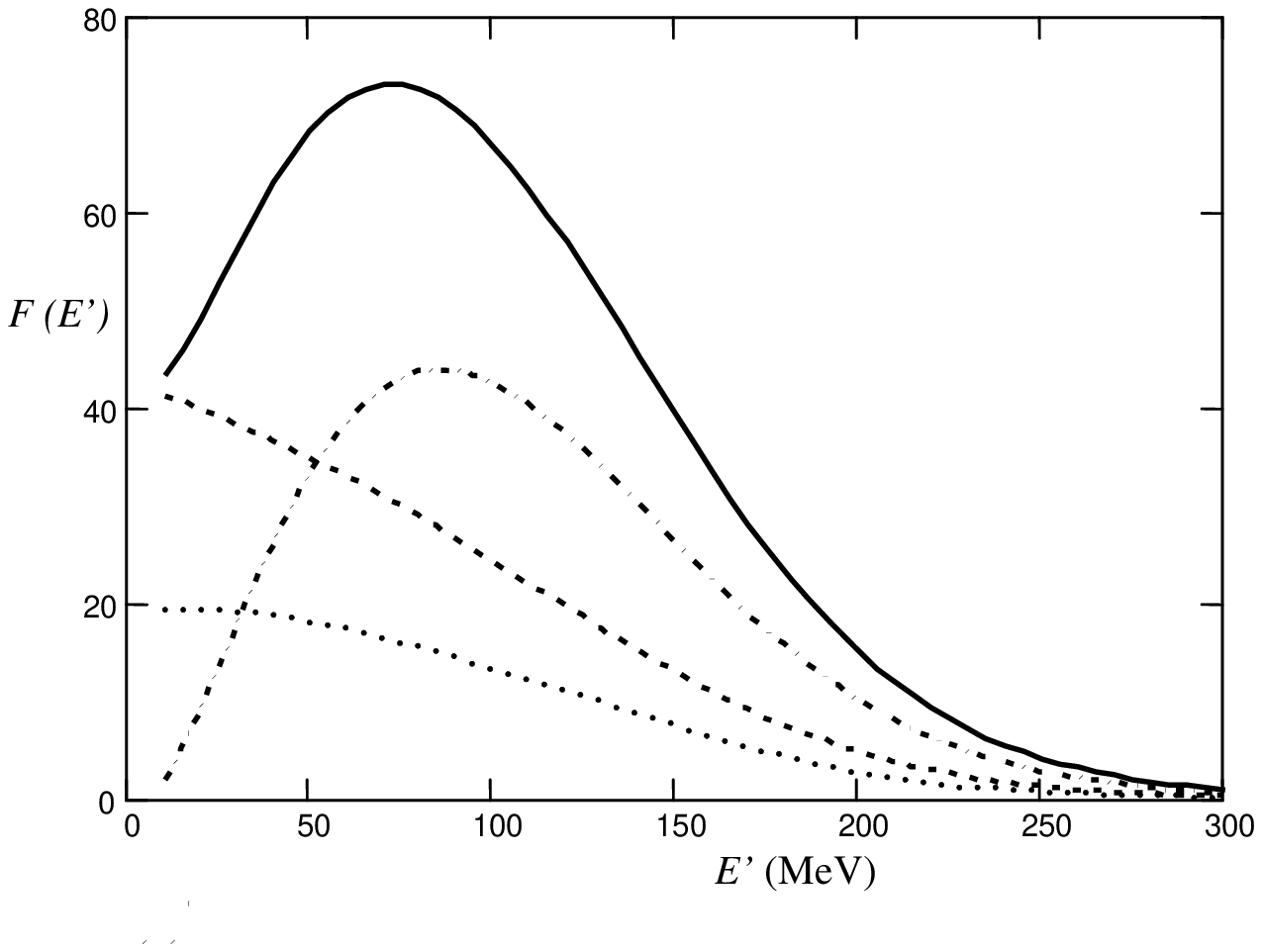}
\caption{The function $F (E ')$ defining the electron contribution 
(dashed line), the proton contribution (dash-dotted line) into the 
$\nu_R$ production rate, and the total rate (solid line)
for the plasma temperature $T =$30 MeV. 
The dotted line shows the result by A. Ayala et al.~\cite{Ayala:2000}.}
\label{fig:rate_function}
\end{figure}

The details of calculations of the rate of creation of the right-handed 
neutrino will be published elsewhere. 
The results of our numerical analysis of the separate contributions of 
the neutrino scattering off electrons and protons, as well as the total 
$\nu_R$ production rate in the typical conditions of the supernova core 
are presented in Fig.~\ref{fig:rate_function}. 

The plotted function $F (E ')$ is defined by the expression
\begin{eqnarray}
\Gamma_{\nu_R} (E ') = \frac{\mu_\nu^2\ T^3}{32\ \pi}\, F (E ')\,.
\label{eq:F(E)_def}
\end{eqnarray}
It is seen from the Fig.~\ref{fig:rate_function} that the proton 
contribution is essential indeed. 
For comparison, the result by A. Ayala et al.~\cite{Ayala:2000} is also 
shown in Fig.~\ref{fig:rate_function}, illustrating a strong underestimation 
of the neutrino chirality flip rate made by those authors. 

The supernova core luminosity for $\nu_R$ emission can be computed as
\begin{eqnarray}
Q_{\nu_R} = V\, \int \frac{\D^3 p '}{(2 \pi)^3} \; E ' \, 
\Gamma_{\nu_R} (E ')\,,
\label{eq:Q_def}
\end{eqnarray}
where $V$ is the plasma volume. 

For the same supernova core conditions as in the 
papers~\cite{Ayala:1999,Ayala:2000}
(plasma volume $V \sim 8 \times 10^{18} 
{\rm cm}^3$, temperature range $T = 30 - 60$ MeV, 
electron chemical potential range $\mu_e = 280 - 307$ MeV), 
we found
\begin{eqnarray}
Q_{\nu_R} = \left(\frac{\mu_\nu}{\mu_{\rm B}}\right)^2 (0.76 - 4.4) 
\times 10^{77} \; \mbox{ergs/s}\,.
\label{eq:Q_res}
\end{eqnarray}

Assuming that $Q_{\nu_R} < 10^{53}$ ergs/s, we obtain 
the upper limit on the neutrino magnetic moment 
\begin{eqnarray}
\mu_\nu < (0.5 - 1.1) \, \times 10^{-12} \, \mu_{\rm B}\,.
\label{eq:mu_fr_Q}
\end{eqnarray}
%


\section{Left-handed neutrino washing away} 

\begin{figure}[htb]
\centering
\includegraphics[width=0.78\textwidth]{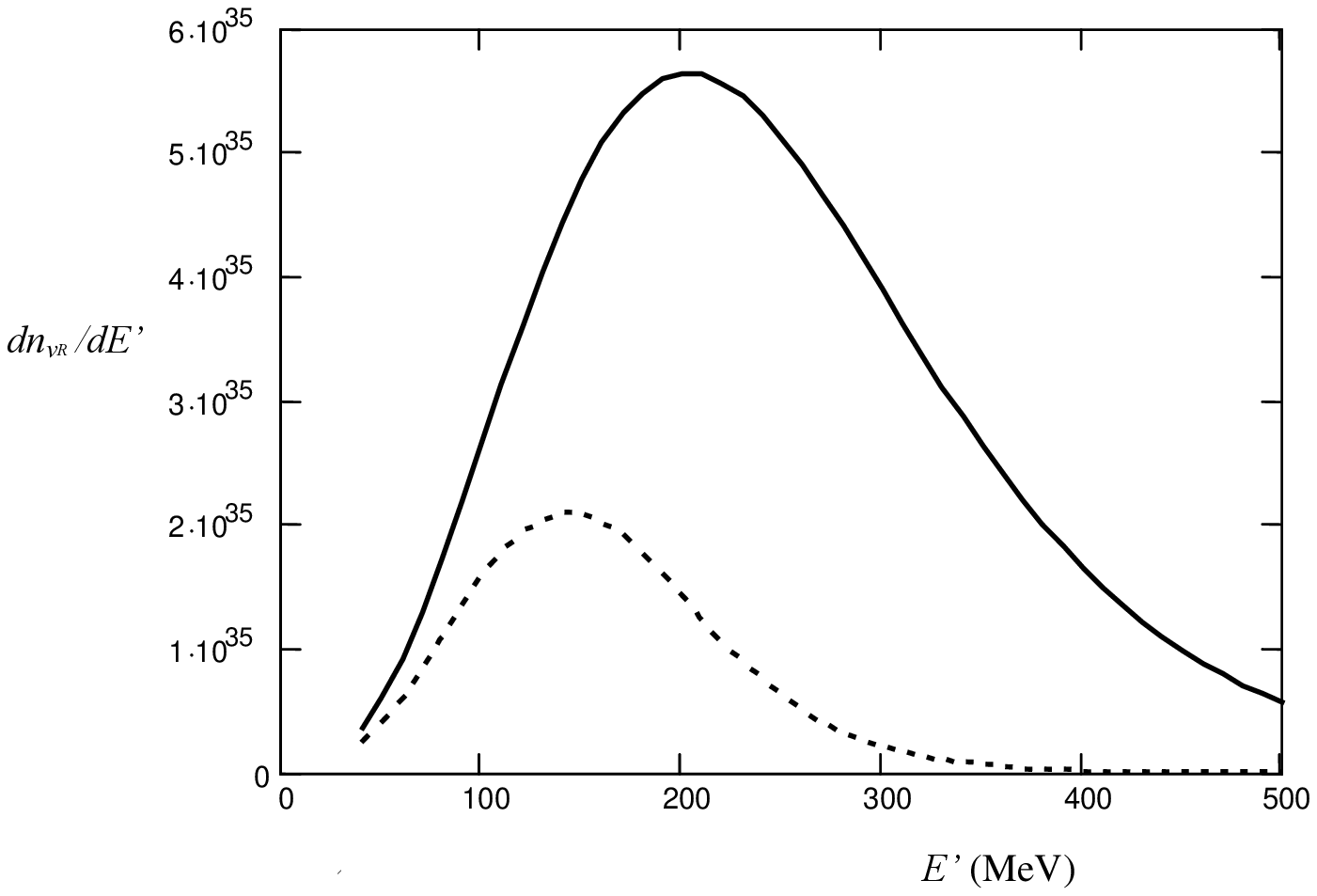}
\caption{The number of right-handed neutrinos 
(for $\mu_\nu = 10^{-12} \, \mu_{\rm B}$) 
emitted per 1 cm$^3$ 
per 1 sec per 1 MeV of the energy spectrum for 
the plasma temperature $T =$60 MeV (solid line) and for 
$T =$30 MeV (dashed line).}
\label{fig:number}
\end{figure}

An additional method can be used to put a bound on the neutrino magnetic 
moment.
Together with the supernova core luminosity $Q_{\nu_R}$, a number of 
right-handed 
neutrinos emitted per 1 sec per 1 cm$^3$ can be defined via the rate 
$\Gamma_{\nu_R} (E ')$ as
\begin{eqnarray}
n_{\nu_R} = \int \frac{\D^3 p '}{(2 \pi)^3} \; \Gamma_{\nu_R} (E ')\,.
\label{eq:n_def}
\end{eqnarray}
The right-handed neutrino energy spectrum, i.e. a number of right-handed 
neutrinos emitted per 1 sec per 1 MeV from the unit volume,  
$\D n_{\nu_R}/\D E '$, 
can be also evaluated numerically. In the Fig.~\ref{fig:number}
we show, taking for definitness $\mu_\nu = 10^{-12} \, \mu_{\rm B}$, 
the result of this calculation for two values of the plasma temperature. 

Integrating the value $\D n_{\nu_R}/\D E '$ over all energies, one obtains 
the number of right-handed neutrinos emitted per 1 cm$^3$ 
per 1 sec. Dividing this to the initial left-handed neutrino 
number density $n_{\nu_L}$ , one can estimate the averaged time of the 
left-handed neutrino washing away, i.e. of the total 
conversion of left-handed neutrinos to right-handed neutrinos. 
For the temperature range $T = 30 - 60$ MeV, and 
for the electron chemical potential $\mu_e \sim 300$ MeV, 
we obtain 
\begin{eqnarray}
\tau \simeq \left(\frac{10^{-12}\,\mu_{\rm B}}{\mu_\nu}\right)^2 (0.14 - 0.36) 
 \; \mbox{sec}\,.
\label{eq:tau}
\end{eqnarray}
In order not to spoil the Kelvin---Helmholtz stage of the protoneutron star 
cooling ($\sim$ 10 sec), this averaged time of the neutrino spin-flip 
should exceed a few seconds. 
Taking the conservative limit $\tau >$ 1 sec, we obtain the bound 
on the neutrino magnetic moment:
\begin{eqnarray}
\mu_\nu < (0.4 - 0.6) \, \times 10^{-12} \, \mu_{\rm B}\,.
\label{eq:mu_fr_num}
\end{eqnarray}

By this means, we improve the best astrophysical upper bound  
on the neutrino magnetic moment by A. Ayala et al.~\cite{Ayala:1999}.
by the factor of 3 to 7.


\section{Conclusions}


\begin{itemize}
\item
We have investigated in detail the neutrino chirality-flip process under the 
conditions of the supernova core. The plasma polarization effects caused both by 
electrons and protons were taken into account in the photon propagator. 
The rate $\Gamma_{\nu_R} (E ')$ of creation of the right-handed 
neutrino with the fixed energy $E '$, the energy 
spectrum, and the luminosity have been calculated. 
\item
From the limit on the supernova core luminosity for $\nu_R$ emission, 
we have obtained the upper bound on the neutrino magnetic moment 
$
\mu_\nu < (0.5 - 1.1) \, \times 10^{-12} \, \mu_{\rm B}\,.
$ 
\item
From the limit on the averaged time of the neutrino spin-flip,
we have obtained the upper bound
$
\mu_\nu < (0.4 - 0.6) \, \times 10^{-12} \, \mu_{\rm B}\,.
$ 
\item
We have improved the best astrophysical upper bound on the neutrino 
magnetic moment by the factor of 3 to 7.
\end{itemize}


\section*{Acknowledgements}


A. K. expresses his deep gratitude to the organizers of the 
XIV-th International Baksan School ``Particles and Cosmology'' 
for warm hospitality.

The work was supported in part 
by the Council on Grants by the President of Russian Federation 
for the Support of Young Russian Scientists and Leading Scientific Schools of 
Russian Federation under the Grant No. NSh-6376.2006.2 
and
by the Russian Foundation for Basic Research under the Grant No. 07-02-00285-a. 



\end{document}